\documentclass[conference]{IEEEtran}
\IEEEoverridecommandlockouts
\usepackage{cite}
\usepackage{amsmath,amssymb,amsfonts}
\usepackage{algorithmic}
\usepackage{graphicx}
\usepackage{textcomp}
\usepackage{xcolor}
\usepackage{threeparttable}
\usepackage{booktabs}
\usepackage[colorlinks,linkcolor=blue,anchorcolor=blue,citecolor=blue]{hyperref}
\def\BibTeX{{\rm B\kern-.05em{\sc i\kern-.025em b}\kern-.08em
    T\kern-.1667em\lower.7ex\hbox{E}\kern-.125emX}}
\definecolor{mygray}{gray}{.9}
\begin{document}

\title{A Multi-Modal Transformer-based Code Summarization Approach for Smart Contracts}

\author{
\IEEEauthorblockN{Zhen Yang\IEEEauthorrefmark{2},
Jacky Keung\IEEEauthorrefmark{2}, Xiao Yu\IEEEauthorrefmark{3}\IEEEauthorrefmark{1}\thanks{*Corresponding author.},
Xiaodong Gu\IEEEauthorrefmark{4}, Zhengyuan Wei\IEEEauthorrefmark{2},\\
Xiaoxue Ma\IEEEauthorrefmark{2}, and Miao Zhang\IEEEauthorrefmark{2}}
\IEEEauthorblockA{
\IEEEauthorrefmark{2}Department of Computer Science, City University of Hong Kong, Hong Kong, China, \\ \{zhyang8-c, zywei4-c, xiaoxuema3-c, miazhang9-c\}@my.cityu.edu.hk, Jacky.Keung@cityu.edu.hk\\
\IEEEauthorrefmark{3} School of Computer Science and Technology, Wuhan University of Technology, Wuhan, China, xiaoyu@whut.edu.cn\\
\IEEEauthorrefmark{4}School of Software, Shanghai Jiao Tong University, Shanghai, China, xiaodong.gu@sjtu.edu.cn}}

\maketitle

\begin{abstract}
Code comment has been an important part of computer programs, greatly facilitating the understanding and maintenance of source code. However,  high-quality code comments are often unavailable in smart contracts, the increasingly popular programs that run on the blockchain.  In this paper, we propose a Multi-Modal Transformer-based (MMTrans) code summarization approach for smart contracts. 
Specifically, the MMTrans learns the representation of source code from the two heterogeneous modalities of the Abstract Syntax Tree (AST), i.e., Structure-based Traversal (SBT) sequences and graphs. 
The SBT sequence provides the global semantic information of AST, while the graph convolution focuses on the local details. The MMTrans uses two encoders to extract both global and local semantic information from the two modalities respectively, and then uses a joint decoder to generate code comments. Both the encoders and the decoder employ the multi-head attention structure of the Transformer to enhance the ability to capture the long-range dependencies between code tokens.
We build a dataset with over 300K $<$method, comment$>$ pairs of smart contracts, and evaluate the MMTrans on it. The experimental results demonstrate that the MMTrans outperforms the state-of-the-art baselines in terms of four evaluation metrics by a substantial margin, and can generate higher quality comments.  
\end{abstract}


\section{Introduction}
\label{introduction}
Automatic code summarization, which generates a brief natural language description for source code, can greatly facilitate programmers in code comprehension and maintenance. Many approaches have been proposed to generate  comments for some common programming languages, such as Java and Python.
In recent years, smart contracts, as programs automatically running on the blockchain \cite{roscheisen1998stanford,tapscott2016blockchain,savelyev2017contract}, have been applied in various business areas to enable more efficient and trustable transactions \cite{gao2020checking,sun2020formal}. More and more developers also devote themselves to the development of smart contracts, and contribute their code to various smart contract communities (e.g., Etherscan.io\cite{Ethereum96:online}). 

However, as an increasingly universal and vital field, automatic code summarization for smart contracts has not gained much attention yet. This may cause a few vital issues:
(1) We have observed that most of smart contract comments are unavailable, thus resulting in great difficulties in comprehending and learning code between developers. (2) Code clone and duplicates in smart contracts is a more common phenomenon than other softwares \cite{gao2020checking, yang2020smart}. 
He \textit{et al.} \cite{he2020characterizing} found that about 10\% of the vulnerabilities were introduced by code clone, 
and the misuse of uncommented code is a principal reason. To this end, it is necessary to automatically generate high-quality code comments for smart contracts.
The challenges of automatic code summariztion usually include:

\textbf{(1) How to extract the semantic information of source code.}

Early researchers, such as Iyer \textit{et al.} \cite{iyer2016summarizing} and Loyola \textit{et al}. \cite{loyola2017neural}, used the plain source code as the input of the code summarization model, which ignored the structural information
of source code. Therefore, most recent works firstly parsed source code to the Abstract Syntax Tree (AST), and then extracted the semantic information of code from AST. For example, Hu \textit{et al}. \cite{hu2020deep} firstly traversed the AST by the (Structure-based Traversal) SBT method to obtain the SBT sequences, then used the plain source code and the SBT sequences as inputs to learn the semantic information of source code. LeClair \textit{et al}. \cite{10.1145/3387904.3389268} regarded the AST as a graph, and used the Graph Neural Network (GNN)-based encoder to model the AST, combined with the RNN-based encoder to model the plain source code. However, ASTs can be represented as multiple modalities, such as SBT sequences and graphs, each focusing on a distinct aspect of the semantic information. Therefore, it is not comprehensive to use a single AST modality to represent the semantic information of code.

\textbf{(2) How to capture the long-range dependencies between code tokens.}

The source code of smart contract methods can be very long. In our experiment dataset, smart contract methods contain 58.72 code tokens on average, and the longest one contains 3,272 code tokens. Previous works often applied the Recurrent Neural Network (RNN), Long Short Term Memory (LSTM), or Gated Recurrent Unit (GRU) models to extract features from their inputs, which have shown to be difficult to capture the long-range dependencies between code tokens  \cite{vaswani2017attention, ahmad-etal-2020-transformer}.

In order to address the above challenges, in this paper, we propose a Multi-Modal Transformer-based (MMTrans) code summarization approach for smart contracts. Firstly, the MMTrans learns the semantic information of source code from the two modalities of the AST, i.e., the SBT sequence and the graph. Specifically, the SBT sequence is globally parsed from the AST using the SBT method, which involves the global semantic information of source code. Meanwhile, MMTrans regards the AST as a graph, and employs the Graph Convolutional Neural Network (GCN) to learn representations of nodes based on their neighboring nodes, thus obtaining the local semantic information of source code. Then, the MMTrans uses a dual-encoder architecture: a SBT encoder for encoding SBT sequences to extract the \textbf{global} semantic information, 
and a graph encoder for encoding graphs to extract the \textbf{local} semantic information.
Finally, the MMTrans uses a joint decoder to decode the outputs of the two encoders and previous generated words to produce the each time step's prediction. In addition, both the encoders and the decoder employ the multi-head attention structure of the Transformer to reinforce the capability of capturing the long-range dependencies between code tokens.

To evaluate the MMTrans, we carefully collect 347,410 $<$method, comment$>$ pairs from 40,932 smart contracts on Etherscan.io \cite{Ethereum96:online}, one of the most popular and active smart contract communities.
We extensively assess the performance of the MMTrans against the three recently proposed approaches (i.e., Hybrid-DeepCom \cite{hu2020deep}, code+gnn+GRU \cite{10.1145/3387904.3389268} , and Vanilla-Transformer \cite{ahmad-etal-2020-transformer}) on the dataset in terms of sentence-level BLEU (S-BLEU), corpus-level BLEU (C-BLEU), ROUGE-LCS F1 and METEOR. 
The experimental results show that the MMTrans performs better than the baselines by 17.23\%-58.45\% in terms of S-BLEU, by 20.74\%-62.49\% in terms of C-BLEU, by 6.78\%-55.60\% in terms of ROUGE-LCS F1, and by 10.17\%-45.50\% in terms of METEOR. We further conduct two groups of ablation experiments to explore the strength of MMTrans. Both the quantitative and instance analysis demonstrate that the MMTrans can generate higher-quality comments for smart contracts.

The main contributions of this paper can be summarized as follows:
\begin{itemize}

    \item We propose a novel Multi-Modal Transformer-based (MMTrans) code summarization approach for smart contracts, which can extract both the global and local semantic information of code, and capture the long-range dependencies between code tokens to generate higher-quality comments.
    \item We build a dataset with totally 347,410 $<$method, comment$>$ pairs for the field of smart contract code summarization. To the best of our knowledge, it is the first large-scale dataset for this task. 
    \item  We open source our replication package, including the dataset \cite{SmartCon45:online} and the source code \cite{yz10191112:online} of the MMTrans for follow-up works.
\end{itemize}

The remainder of this paper is organized as follows. Section \ref{relatedworkandbackground} presents the related work and background. Section \ref{APPROACH} and Section \ref{Data Preparation} elaborate on the proposed approach and data preparation. Section \ref{experimental setup} and Section \ref{experimental results} discuss the experiment design and results. Section \ref{threats} demonstrates the threats to validity. Finally, Section \ref{conclusion} concludes the work of this paper.

\section{RELATED WORK AND BACKGROUND}
\label{relatedworkandbackground}
\subsection{Code Summarization}
\label{related work}
Automatic code summarization approaches can be divided into two main categories, i.e., heuristic/template-driven approaches and AI/data-driven approaches \cite{leclair2019neural}. Haiduc \textit{et al}. \cite{2010On,haiduc2010supporting} for the first time coined the term “source code summarization”, and proposed a heuristic-based approach, which applied text retrieval techniques to select some important keywords as the generated code comments.  Following Haiduc \textit{et al}.'s works, many researchers \cite{eddy2013evaluating,haiduc2010supporting,sridhara2010towards,sridhara2011automatically,moreno2013automatic,wong2015clocom,mcburney2015automatic,rodeghero2015eye} proposed to design a set of heuristic rules or create some manually-crafted templates to generate code comments. 
Due to the rapid development of deep learning technology, the recently proposed code summarization approaches are almost AI/data-driven approaches.  Iyer \textit{et al.} \cite{iyer2016summarizing} for the first time proposed an AI/data-driven code summarization approach, which used the LSTM networks with attention to generate descriptions for C\# code snippets and SQL queries.  The subsequent works \cite{loyola2017neural,lu2017learning,liang2018automatic,hu2018deep,wan2018improving,leclair2019neural,DBLP:conf/iclr/AlonBLY19,fernandes2018structured,wang-etal-2019-learning,hu2020deep,10.1145/3387904.3389268,zhang2020retrieval} almost adopted the Sequence-to-Sequence (Seq2Seq) model with attention mechanism. The major difference of the approaches are the input to the Seq2Seq model. For example, Loyola \textit{et al.} \cite{loyola2017neural} input the plain source code into the Seq2Seq model, Lu \textit{et al.} \cite{lu2017learning} used the API sequences as the input, and Fernandes \textit{et al.} \cite{fernandes2018structured} used the graph representations of source code as the input. 
In addition, in order to extract more information from the source code, Hu \textit{et al}. \cite{hu2018summarizing, hu2020deep} and LeClair \textit{et al}. \cite{leclair2019neural,10.1145/3387904.3389268} proposed some multi-input Seq2Seq models. For example, Hu \textit{et al}. \cite{hu2020deep} input the plain source code and the SBT sequences to the Seq2Seq model, in order to learn both the lexical and structural information from the source code and the AST. LeClair \textit{et al}. \cite{10.1145/3387904.3389268} used the graph representation of the AST and plain source code as the inputs. However, it is not comprehensive that these works \cite{hu2018deep,hu2020deep,10.1145/3387904.3389268} used single AST modality to represent the semantic information of code. In addition, the proposed Seq2Seq models \cite{iyer2016summarizing,loyola2017neural,lu2017learning,liang2018automatic,hu2018deep,wan2018improving,leclair2019neural,DBLP:conf/iclr/AlonBLY19,fernandes2018structured,wang-etal-2019-learning,hu2020deep,10.1145/3387904.3389268} mainly applied RNN, LSTM, or GRU to extract the code feature, which may fail to capture the long-range dependencies between code tokens. Therefore, Ahmad \textit{et al}. \cite{ahmad-etal-2020-transformer} empirically investigated the advantage of using the Transformer model for the source code summarization task. However, they only adopted the plain source code as the single input, thus ignoring the structure information of source code. 

\label{background}

\subsection{Structure-based Traversal}
\label{Structure-based Traversal}
Hu \textit{et al}. \cite{hu2018deep} proposed the Structure-based Traversal (SBT) method, which converts the ASTs into specially formatted sequences by globally traversing the ASTs. Specifically, it applies the ``type'' and ``value'' of nodes to represent the structural and lexical information of code, respectively, and adopts a series of brackets to keep the AST structure to ensure the generated sequence is recoverable to the original AST. The SBT method was applied in some previous code summarization models, such as ast-attendgru \cite{leclair2019neural}, Dual Model \cite{wei2019code}, and Hybrid-DeepCom \cite{hu2020deep}, and was proved its powerful ability in preserving code structural and lexical information. Therefore, we regard the SBT sequence as the one modality of AST, and adopt the SBT method to represent the global semantic (i.e., both structural and lexical) information as an input of the MMTrans. Taking the smart contract snippet in the Figure \ref{Fig.1} as an example, the method named $\_tokensToSell$ is firstly transformed to its AST format, and then the SBT sequence is further extracted from the AST. The non-leaf nodes are represented by their ``type'' (such as ``FunctionDefinition'' and ``Block'' in bold). For leaf nodes, they are represented by the format of ``type\_value'' (such as ``Visibility\_private'', etc.) in the original paper \cite{hu2018deep}. However, in our experiment, based on the original SBT sequences, we split the ``type'' and ``value'' for leaf nodes, and further split the camelCase and snake\_case tokens of leaf nodes' ``values'' (such as from ``\_tokensToSell'' to ``\_tokens'', ``To'', and ``Sell'' in italic) to reduce the Out-of-Vocabulary (OOV) tokens, the detailed data processing methods are elaborated in Section \ref{Preprocessing}. 

\subsection{Graph Convolutional Neural Network}
\label{Graph Convolutional Neural Network}
\begin{figure*}[htbp] 
\centering 
\includegraphics[width=1\textwidth]
{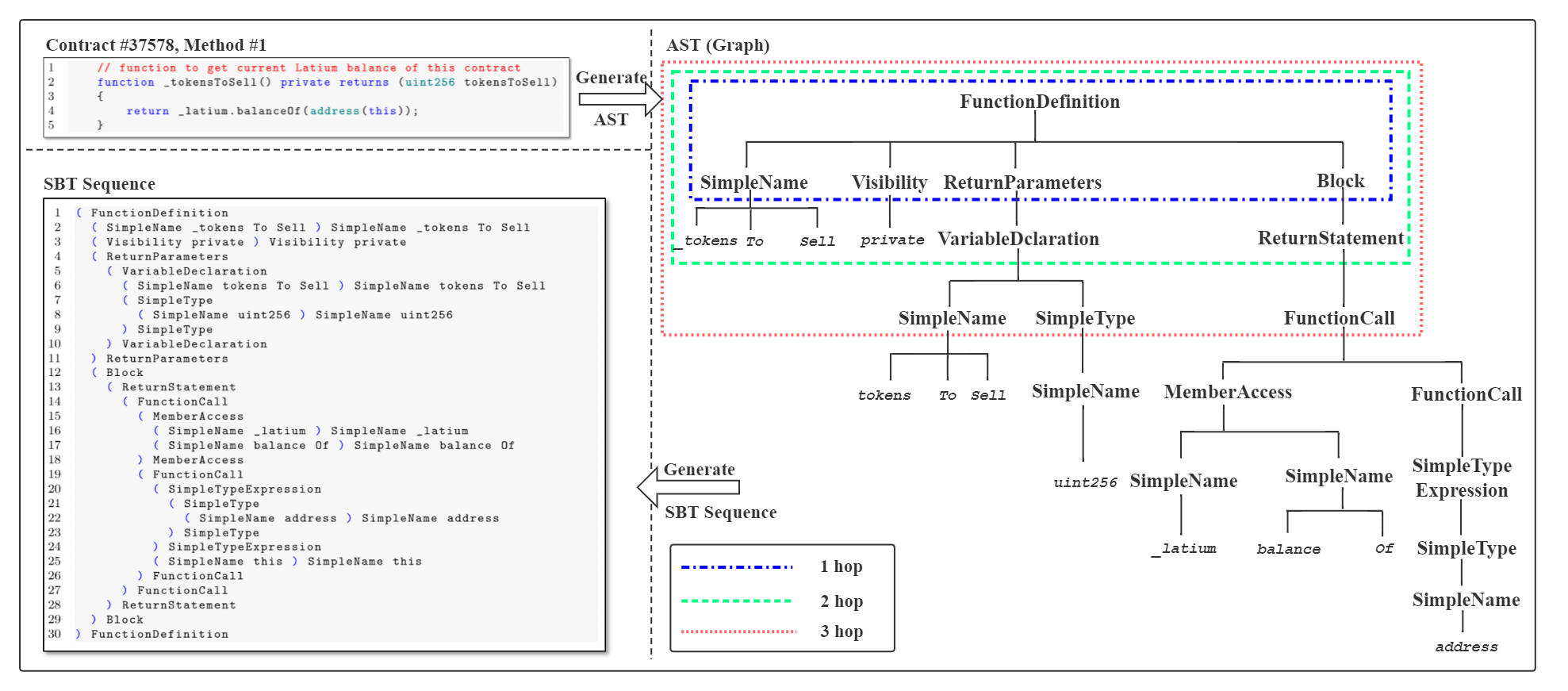} 
\caption{The AST (Graph) and SBT Sequence of the Smart Contract Method named \_tokensToSell} 
\label{Fig.1} 
\end{figure*}

The Graph Convolutional Neural Network (GCN) is designed for information propagation along the edges between nodes, where the hop, representing the layers of GCN, is a critical hyper-parameter. With the hop increasing, each node can aggregate a larger range of information from its neighbors, thereby focusing on a wider scope of local semantic information. Since the node embeddings in GCN includes both the ``type'' and ``value'' in ASTs, the GCN implies both the lexical and structural semantics of the AST integrated along the edges. Previous code summarization works, such as graph2seq \cite{xu2018graph2seq} and code+gnn+GRU \cite{10.1145/3387904.3389268}, also have proved the strength of GCN on locally distilling the semantic information from the AST and achieved promising results. Therefore, we regard the graph as another modality of AST, and adopt it as a parallel input of the MMTrans relative to the SBT sequences. The AST (graph) in Figure \ref{Fig.1} is an example generated from the method named $\_tokensToSell$, and shows the hop of 1,2 and 3 of the root node (i.e., ``FunctionDefinition''). Intuitively, it can aggregate the information from its neighbors of ``SimpleName'', ``Visibility'', ``ReturnParameters'' and ``Block'' by the convolution of the first hop; and it can also indirectly aggregate the more extensive information from its children nodes' neighbors by increasing the hop number. More formally, the graph convolution process can be defined by the following layer-wise propagation rule:
\begin{equation}
    H^{(l+1)} = \sigma(\tilde{A}H^{(l)}W^{(l)}) \label{eq1},
\end{equation}
where the $H^{l}$ is the nodes embedding matrix at the layer $l$, the $\tilde{A} = A + I_{N}$ is the adjacency matrix $A$ of a particular graph with added self-connection, $I_{N}$ is the identity matrix, $W^{l}$ is a layer-specific trainable weight matrix, and $\sigma$ is the activation function \cite{kipf2016semi}.

\subsection{Transformer-related Structures}
\label{Transformer-related Structures}
\subsubsection{Positional Encoding}
Since the multi-head attention is not the recurrent structure, it needs the positional encoding to inject order information into the token embedding vectors. In this work, we follow one of the positional encoding approaches proposed by Vaswani \textit{et al.} \cite{vaswani2017attention}, which defines the specific pattern that model learns. This kind of positional encoding rule can be defined by the equations \ref{eq2} and \ref{eq3}, where $pos$ is the token position in a sequence, $i$ is the dimension index, and $d$ is total dimension of the token embedding vector. 
\begin{align}
\Large{PE_{(pos, 2i)} = sin(pos / 10000^{2i / d})} \label{eq2}\\
\Large{PE_{(pos, 2i+1)} = cos(pos / 10000^{2i / d})} \label{eq3}
\end{align}

\subsubsection{Multi-head Attention}
\label{Multi-head Attention}
In order to pay attention from different perspectives and capture the long-range dependencies in sequences, Vaswani \textit{et al.} \cite{vaswani2017attention} also introduced the multi-head attention mechanism. The details are given by the following equations:
\begin{equation}
q_{1}, ..., q_{J} = split(QW^{Q}) \label{eq4}
\end{equation}
\begin{equation}
k_{1}, ..., k_{J} = split(KW^{K}) \label{eq5}\\
\end{equation}
\begin{equation}
v_{1}, ..., v_{J} = split(VW^{V}) \label{eq6}\\
\end{equation}
\begin{equation}
head_{j} = Softmax(\frac{q_{j}k_{j}^{T}}{\sqrt{d_{k}}})v_{j},\ j = 1,...,J\label{eq7}\\
\end{equation}
\begin{equation}
MultiHead(Q, K, V) = Concat(head_{1}, ..., head_{J})W^{o}\label{eq8}
\end{equation}
Here, the $Q\in \mathbb{R}^{Q_{l}\times Q_{d}}$, $K\in \mathbb{R}^{K_{l}\times K_{d}}$ and $V\in \mathbb{R}^{K_{l}\times V_{d}}$ represent the matrices of query, key and value, respectively, while $q_{j}\in \mathbb{R}^{Q_{l}\times q_{d}}$, $k_{j}\in \mathbb{R}^{K_{l}\times k_{d}}$, $v_{j}\in \mathbb{R}^{K_{l}\times v_{d}}$ represent their splitted matrices for $head_{j}$. Specifically, $q_{d} = k_{d} = v_{d} = d_{model}/J$. The $W^{Q} \in \mathbb{R}^{Q_{d}\times d_{model}}$, $W^{K} \in \mathbb{R}^{K_{d}\times d_{model}}$, $W^{V} \in \mathbb{R}^{V_{d}\times d_{model}}$ are the three trainable weight matrices. The equation \ref{eq7} describes the Scaled Dot-Product Attention output of $head_{j}$, where the $d_{k}$ is the scaling factor equals to $k_{d}$. Finally, after the concatenating from all heads and the linear transformation with $W^{o} \in \mathbb{R}^{Jv_{d}\times d_{model}}$, we obtain the output of the multi-head attention in equation \ref{eq8}\cite{vaswani2017attention}.

\subsubsection{Point-Wise Feed-Forward Networks}
\label{Point-Wise Feed-Forward Networks}
This is another module of Transformer that applied in \cite{vaswani2017attention}. It is composed of two dense layers with a ReLU activation function in between, which can be defined by the equation \ref{eq9}, where $W_{1}$ and $W_{2}$ are the weight matrices of each layer, $b_{1}$ and $b_{2}$ are their corresponding bias, and $x$ is the input matrix. The dimensionality of the inner-layer $d_{ff}$ is a hyper-parameter in \cite{vaswani2017attention}.
\begin{equation}
    FFN(x) = max(0, xW_{1} + b_{1})W_{2} + b_{2} \label{eq9}
\end{equation}

\begin{figure*}[htbp] 
\centering 
\includegraphics[width=1\textwidth]
{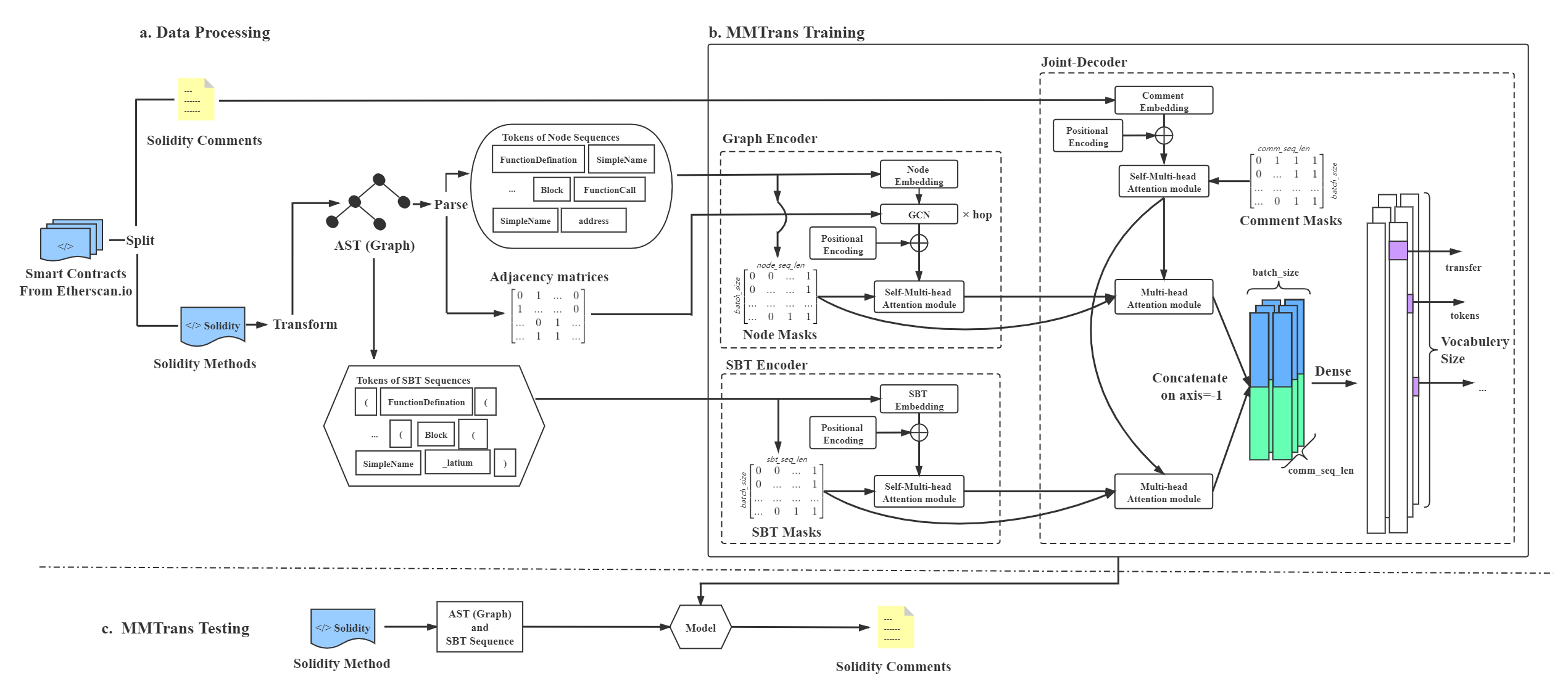} 
\caption{The Overall Framework of MMTrans} 
\label{Fig.2} 
\end{figure*}

\begin{figure}[htbp]
\centering
\includegraphics[width=0.5\textwidth]
{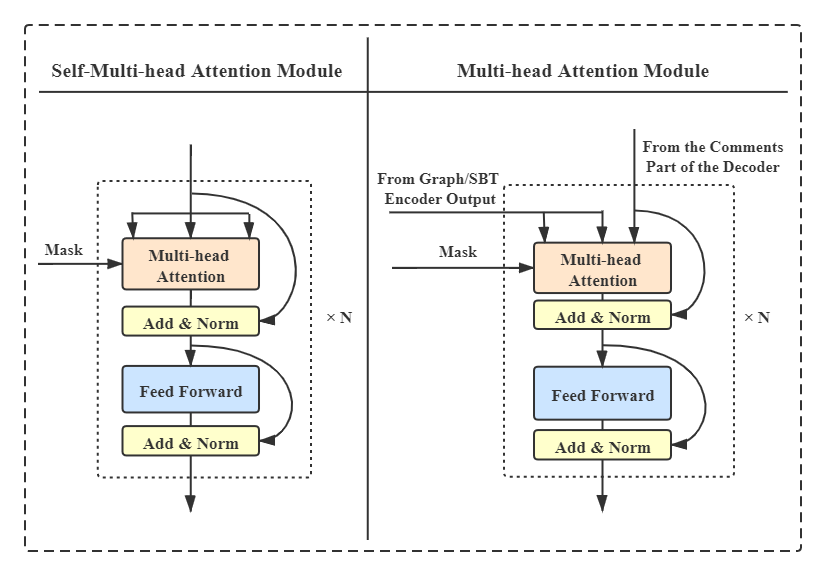}
\caption{(Self) Multi-head Attention Module} 
\label{Fig.3}
\end{figure}

\section{APPROACH}
\label{APPROACH}
The whole framework of the MMTrans illustrated in Figure \ref{Fig.2} includes the three stages: the data processing, the MMTrans training, and the MMTrans testing. The source code we obtained from the Etherscan.io is parsed and processed into a parallel corpus of smart contract methods and their corresponding comments. To comprehensively learn the semantic information of source code, we transform the smart contract methods to graphs (i.e., ASTs) and SBT sequences, respectively, as the inputs of the MMTrans. The MMTrans consists of the two encoders (i.e., the graph encoder and SBT encoder) and a joint decoder. The node sequences and edges (i.e., their corresponding adjacency matrices) of graphs are fed into the graph encoder to learn the local semantic information, while the SBT sequences are fed into the SBT encoder to learn the global semantic information. Subsequently, in the training stage, the joint decoder integrates comment sequences, the graph encoder outputs, and the SBT encoder outputs to produce a batch of sentences under the teacher forcing method. Finally, the back-propagation is executed based on the predefined loss function to optimize the whole network. However, in the testing stage, the joint decoder integrates the previously generated comment words, and above two encoder outputs to predict one word at each time step. It is noticeable that we do not include plain source code as one part of inputs for the MMTrans to learn the lexical information of source code, because each of the modality of AST we adopt already contains both the lexical and structural information, as mentioned in \ref{Structure-based Traversal} and \ref{Graph Convolutional Neural Network}.

 

\subsection{Graph Encoder}
Initially, for a batch of node sequences $X\in \mathbb{R}^{N_{batch}\times l}$, where $l$ represents the maximum length of node sequences of this batch and $N_{batch}$ represents the batch size, the graph encoder firstly embeds the node sequences $X$ with the embedding size $d = 256$. Then, the GCN layer takes the embedding layer output and the edges $E \in \mathbb{R}^{N_{batch}\times l \times l}$ as the inputs to perform the graph convolution that we described in the Section \ref{Graph Convolutional Neural Network}. Previous work by LeClair \textit{et al.} \cite{10.1145/3387904.3389268} has proved that the graph convolution layers $hop=2$ is the best setting for the code summarization task, here we follow their setup and fix the $hop=2$. Each node at the end of the GCN layer aggregates the neighboring information after the graph convolution. Since we adopt the pre-order traversal method to produce the node sequences from ASTs, the node sequences imply the original appearance order of the tokens in the source code, which is also the necessary information that needs to be considered.
Therefore, we add the positional encoding matrices $PE\in \mathbb{R}^{N_{batch}\times l\times d}$ to the output of the GCN layer, so as to inject the position-wise information.

Subsequently, the output of the GCN layer is imported into the Self-Multi-head Attention Module (SMAM) to distill their semantic information further. The internal structure of the SMAM is demonstrated in Figure \ref{Fig.3}. It includes the multi-head attention, layer-normalization, and point-wise feed-forward network (the $d_{ff}$ of the network is set to 512), which are the principal parts of the Transformer introduced in Section \ref{Transformer-related Structures}.
Initially, we set the head number $J = 4$, so that the multi-head attention can focus on each node sequence from four different representation subspaces \cite{vaswani2017attention}. Moreover, the $d_{model}$ in the multi-head attention is set to 256, representing the width of this module; while the number of layers $N$, representing the depth of this module, is set to 1. Besides, we also adopt the node mask $M \in \mathbb{R}^{N_{batch}\times l}$ generated from the batch data to avoid distracting attention by $<$PAD$>$ tokens. Finally, we obtain the output of SMAM $\hat{X}\in \mathbb{R}^{N_{batch}\times l\times d_{model}}$ at the end of the graph encoder, and the whole process can be described by the following equation, where the $f$ is the abstract mapping function constructed by the graph encoder:
\begin{equation}
    \hat{X} = f(X,\ E,\ PE,\ M) \label{eq10}
\end{equation}

\subsection{SBT Encoder}
For a batch of SBT sequences $X^{'}\in \mathbb{R}^{N_{batch}\times l^{'}}$, where the $l^{'}$ represents the maximum length of SBT sequences of this batch, the SBT encoder also firstly embeds the $X^{'}$ with the embedding size $d = 256$ and injects the position-wise information with $PE^{'}\in \mathbb{R}^{N_{batch}\times l^{'}\times d}$. Subsequently, the SBT encoder adopts the SMAM to extract the semantic information with the same hyper-parameters and uses the SBT mask $M^{'} \in \mathbb{R}^{N_{batch}\times l^{'}}$ to avoid distraction. Thereby, we obtain the final output of SBT encoder $\hat{X^{'}}\in \mathbb{R}^{N_{batch}\times l^{'}\times d_{model}}$. The equation below illustrates the abstract mapping function $f^{'}$ of the SBT encoder:
\begin{equation}
    \hat{X^{'}} = f^{'}(X^{'},\ PE^{'},\ M^{'}) \label{eq11}
\end{equation}

\subsection{Joint Decoder}
Similarly, for a batch of comment sequences $Y \in \mathbb{R}^{N_{batch}\times l^{Y}}$, where the $l^{Y}$ represents the maximum length of comment sequences of this batch, the joint decoder also firstly embeds the $Y$ with the embedding size $d = 256$, and injects the positional information $PE^{Y}\in \mathbb{R}^{N_{batch}\times l^{Y}\times d}$. Then, the features of comment sequences are extracted by the SMAM with the same hyper-parameters. It should be noticed that the comment mask $M^{Y} \in \mathbb{R}^{N_{batch}\times l^{Y} \times l^{Y}}$ is an addition of a padding mask and a look-ahead mask (i.e., an upper triangular matrix), which is used for avoiding distraction and information leakage of the subsequent tokens in training \cite{vaswani2017attention}. 

Next, two Multi-head Attention Modules (MAMs) are introduced in the joint decoder with the same hyper-parameters to SMAM. The internal structure is shown in Figure \ref{Fig.3}. One for comment sequences and the output of graph encoder $\hat{X}$, another for the comment sequences and the output of SBT encoder $\hat{X^{'}}$, thereby learning which tokens from the two encoders are important to the inference of comments, respectively.
The outputs from the two MAMs are concatenated together on their last axis, representing a merge of their respective prediction for the comments. Finally, the merged output is fed into a linear transformation layer followed by a softmax function to produce the probability distribution over the vocabulary, thereby obtaining the final result $\hat{Y} \in \mathbb{R}^{N_{batch}\times l^{Y} \times S}$ of the MMTrans, where the $S$ represents the comment vocabulary size. The outlined mapping function $f^{Y}$ of joint decoder can be summarized by the equation below:
\begin{equation}
    \hat{Y} = f^{Y}(Y,\ \hat{X},\ \hat{X^{'}},\ PE^{Y},\ M^{Y})\label{eq12}
\end{equation}

Similar to most of the Seq2Seq models, we define the loss function for each batch as equation \ref{eq13}, where the $l^{y}_{i}$ represents the length of the $i$th real comment (ground truth) removed $<$PAD$>$ tags, and $p(\hat{Y}_{ij}^{(z)})$ represents the probability that the $j$th token in the $i$th sample is the $z$th ($z$ is the ground truth) word in the whole comment vocabulary.
\begin{equation}
    Loss_{batch} = -\frac{1}{N_{batch}}\sum_{i = 1}^{N_{batch}}\frac{1}{l^{y}_{i}}\sum_{j = 1}^{l^{y}_{i}}log \ p(\hat{Y}_{ij}^{(z)}) \label{eq13}
\end{equation}

\section{Data Preparation}
\label{Data Preparation}
\subsection{Preprocessing}
\label{Preprocessing}
The raw dataset is provided by Zhuang \textit{et al.} \cite{ijcai2020-454}, which was collected from the Etherscan.io. The dataset contains totally 40,932 Ethereum smart contracts written in solidity with 933,146 \textit{normal methods}, 73,533 \textit{modifiers}, and 12,482 \textit{fallback methods}. \textit{Normal methods} includes \textit{constructors} and other \textit{functional methods}; \textit{Modifiers} are used to change the behaviour of functions in a declarative way, such as checking a condition prior to executing the function; \textit{Fallback methods} will be executed on a call to the contract, if none of the other methods match the given method signature; another kind of method in solidity is the \textit{receive method}, which is also a kind of \textit{fallback method}, and is first introduced in the version of solidity 0.6.0 to receive ether\cite{Solidity48:online}.
According to our preliminary study on the whole dataset, we find 1-to-\textit{n}  matching problem between methods and comments, i.e., the same code may correspond to different comments among most of the \textit{fallback methods}, which will confuse the MMTrans in training. The main reason is that most of the developers use \textit{fallback methods} for reverting. But different \textit{fallback methods} in the different smart contracts will revert different objectives. In addition, generating comments for \textit{constructors} and \textit{receive methods} is trivial, because their functionalities are fixed and easy for machines to learn. Furthermore, there is also no \textit{receive method} in the whole dataset. Therefore, in our experiment, we only consider normal \textit{functional methods} and \textit{modifiers}, and we remove the methods without comments. The remaining data are formulated as $<$method, comment$>$ pairs for further process.

According to the introduction of annotation in the solidity doc \cite{Solidity48:online} and our observation, we find that smart contract developers tend to place their comments under the NatSpec tags by the priority order of $@notice$, $@dev$, $@return$ or just ``//'' and ``/**/''. Therefore, we extract the texts behind the $@notice$ tag firstly, if there is no $@notice$ tag, we extract the texts behind the $@dev$ tag, and so on. In addition, following the prior similar work \cite{hu2020deep, wei2019code, hu2018deep, 10.1145/3387904.3389268}, we also use the first sentence of texts as the ground truth comment, which typically describes the functionality of the particular method. Moreover, we remove those $<$method, comment$>$ pairs whose comments contain less than 4 words for better computation of the BLEU-4 score \cite{hu2020deep}. Finally, we collect the  347,410 $<$method,  comment$>$ pairs from the 40,932 smart contracts.

\subsection{Data Transformation}
We transform the source code to SBT sequences and graphs (represented by xml format) respectively by utilizing the solidity-parser-antlr \cite{federico21:online}. To reduce the Out-Of-Vocabulary (OOV) tokens and facilitate the model to capture token representation, we split the camelCase and snake\_case tokens for the ``value'' of leaf nodes in both SBT sequences and graphs. Further, we parse the graphs to their corresponding node sequences and edges (i.e., adjacency matrices), and formulate the $<$method, comment$>$ pairs to (SBT sequence, node sequence, adjacency matrix, comment) tuples. The statistics of the lengths of the SBT sequences, graph node sequences, and comments are shown in the Table \ref{Statistics for Data Lengths}. The average length of the SBT sequences, graph node sequences and comments are 241.44, 75.92, and 10.04, respectively. Here, we set the maximum length of comments to 20, and remove those tuples with comment length larger than 20 (i.e., retain 96.53\% of the whole data). We also set the maximum length of SBT sequences and graph nodes to 600 and 200, and truncate those excessively long sequences. Afterward, we replace the $numeral$ and $string$ in source code with $<$NUM$>$ and $<$STR$>$ respectively. Since the $address$ of smart contracts is a special object in smart contracts and is composed of fixed 40 hexadecimal digits, we generalize $address$ constants by $<$ADDR$>$. Finally, for the SBT sequences and comments, we add $<$START$>$ and $<$END$>$ tokens to represent the start and end of sequences. However, for graph nodes sequences, we keep them intact, which is also a common operation in GCN \cite{10.1145/3387904.3389268}. 

\begin{table}[htbp]\caption{Statistics for Data Lengths}\label{Statistics for Data Lengths}
\centering
\begin{tabular}{cccccc}
\toprule
\multicolumn{6}{c}{The Lengths of the SBT Sequences}                                  \\ 
\midrule
Avg.   & Mode & Median & $\leq 200$ & $\leq 400$ & $\leq 600$ \\
241.44 & 69   & 163    & 59.49\%        & 85.69\%        & 94.92\%        \\ \bottomrule
\toprule
\multicolumn{6}{c}{The Lengths of the Graph Node Sequences}                                    \\ \midrule
Avg.   & Mode & Median & $\leq 100$ & $\leq 150$ & $\leq 200$ \\
75.92  & 21   & 52     & 76.77\%        & 89.98\%        & 95.27\%        \\ \bottomrule
\toprule
\multicolumn{6}{c}{The Lengths of the Comments}                                       \\ \midrule
Avg.   & Mode & Median & $\leq 10$  & $\leq 20$   & $\leq 30$  \\
10.04  & 8    & 10     & 57.18\%        & 96.53\%        & 99.44\%        \\ \bottomrule
\end{tabular}
\end{table}

\subsection{Generating Vocabularies and Input Pipeline}
The above procedures yield 317,680 tuples. We randomly select 90\% (285,912 samples) of them for training, 5\% (15,884 samples) of them for validation, and 5\% (15,884 samples) of them for testing. We remove duplicated samples in the validation set and the testing set that are already included in the training set to avoid data leakage, and finally remain 1,185 and 1,159 samples, respectively. Then, we generate vocabularies for the SBT sequences, the graph node sequences and the comments, respectively on the training set. The size of above vocabularies are 10441, 10431, and 13174, respectively. Furthermore, we set a $<$UNK$>$ token in each of the above vocabulary to substitute the OOV tokens in the validation set and the testing set. 

To prepare the data input pipeline, we randomly select 100 samples to form each batch without replacement. For each input tunnel (i.e., SBT sequences, graph node sequences, adjacency matrices, and comment sequences) of each batch, we append the special tag $<$PAD$>$ to pad them to the maximum length of their batch.

\section{EXPERIMENT DESIGN}
\label{experimental setup}
In this section, we propose our research questions and the corresponding methodologies. Meanwhile, we elaborate on the baselines from the recent similar works, the evaluation metrics, and the experimental devices.

\subsection{Research Questions}
\label{Research Questions}
Our evaluation aim to verify if the MMTrans adopting the multi-modalities of AST and  the multi-head attention structure outperforms the state-of-the-art baselines, and why  the above two points cause the outperformance of the MMTrans. Based on these concerns, we propose the following Research Questions (RQs):
\begin{itemize}
\item \textbf{$RQ_{1}$:} How effective is the MMTrans compared with the state-of-the-art baselines introduced in Section \ref{baselines}?
\item \textbf{$RQ_{2}$:} How does the head number of the multi-head attention structure affect the performance of the MMTrans?
\item \textbf{$RQ_{3}$:} What are the advantages of using the multi-modalities of AST and the multi-head attention structure?
\end{itemize}

Inspired by the previous works \cite{hu2018deep,hu2020deep,10.1145/3387904.3389268}, we make the first attempt of utilizing multi-modalities of AST (i.e., both SBT sequences and graphs) and the multi-head attention structure to construct the framework of dual encoders along with a joint decoder, thereby proposing the MMTrans. To this end, the $RQ_{1}$ is put forward to evaluate the MMTrans against other state-of-the-art models in terms of the metrics in Section \ref{metrics}.

On the other hand, the head number in the multi-head attention structure has shown to be important in attention allocation from different representation subspaces. 
To this end, we put forward the $RQ_{2}$ in this work to explore how the head number affects the MMTrans learning in smart contracts code summarization.  

Finally, the aim of the $RQ_{3}$ is to investigate why the MMTrans is more effective in generating code comments, and how its heterogeneous multi-modalities of AST and the multi-head attention structure contribute to the generation of higher-quality comments for smart contracts, respectively.

\subsection{Methodology}
\label{Methodology}
To answer the $RQ_{1}$, we reproduced three baselines that highly related to our work. Following LeClair \textit{et al.}'s \cite{10.1145/3387904.3389268} experiment setup, we try to set the hyper-parameters of the baselines consistent with the MMTrans for the fair comparison. The detailed hyper-parameters setting for the MMTrans has been elaborated in Section \ref{APPROACH}. Besides, we adopt the same training strategy for each model during the experiment. Specifically, we set the maximum training epoch to 50. Each model is validated every 500 minibatches on the validation set by sentence-level BLEU. We save the model with the best validation performance, and adopt early stopping with the patience of 5 to avoid the model overfitting and save computation cost. Furthermore, we adopt Adam \cite{kingma2014adam} as the training optimizer, and follow the learning rate decay schedule in \cite{vaswani2017attention}. After the above training operations for each model, we choose their respective best performing model to evaluate on the testing set in terms of the metrics in Section \ref{metrics}, and report the comparison result in this paper.

For the $RQ_{2}$, we fix the other hyper-parameters of the MMTrans, and set the head number $J = \{2, 4, 8, 16, 32\}$ respectively. For each adjustment of $J$, we retrain the MMTrans following the training strategy and report the experimental result in terms of the automated metrics in this paper. The reason for the choice of above head numbers is that they should be divisible by $d_{model}$, which is a necessary prerequisite for the multi-head attention structure.

Finally, for the $RQ_{3}$, we evaluate the strength of the heterogeneous multi-modalities of AST and the multi-head attention structure by ablation experiments separately. (1) Firstly, we construct an incomplete-MMTrans (i-MMTrans) utilizing the graph and plain source code as double inputs. Then, we adopt the i-MMTrans to compare with the third baseline (i.e., vanilla-Transformer with only the plain source code input) and the MMTrans with both SBT sequences and graphs as inputs, thereby exploring the contribution of SBT sequences and graphs, respectively. (2) However, for the evaluation of the strength of the multi-head attention structure, we adopt the comparison between the i-MMTrans and the second baseline, i.e., code+gnn+GRU, because they both adopt the graph and plain source code as double inputs. But the i-MMTrans adopts the multi-head attention structure, while the code+gnn+GRU adopts the GRU-based structure. 

\subsection{Baselines}
\label{baselines}
We compare the MMTrans with the three baselines that are published in last year and directly related to our work. We list their detailed information as below:

\textbf{(1) Hybrid-DeepCom:} Hu \textit{et al.} \cite{hu2020deep} exploited the plain source code and SBT sequence with only nodes' ``type'' as the double inputs to extract structural and lexical information of source code, respectively. They also construct a GRU-based Seq2Seq model to process the double inputs and generate comments of Java methods. This approach was published in the EMSE volume 25, 2020, and is the latest representative approach that adopts SBT sequences as input separately with the plain source code.

\textbf{(2) code+gnn+GRU:} LeClair \textit{et al.} \cite{leclair2019neural} proposed to utilize the ASTs (graphs) and the plain source code as the double inputs to improve the code summarization performance based on their previous work \cite{leclair2019neural} published in the ICSE'2019. This approach was presented in the ICPC'2020 and is the latest work adopting GCN in the code summarization task.

\textbf{(3) Vanilla-Transformer:} Ahmad \textit{et al.} \cite{ahmad-etal-2020-transformer} proposed to adopt the Transformer to solve the Java source code comments generation task, which is the first attempt of utilizing the Transformer in this field. This work was published in the ACL'2020. We adopt their base model (i.e., Vanilla-Transformer) to involve in the comparison, rather than their full model with the relative positional encoding and copy attention, because these tricks make it difficult to distinguish the improvement of multi-modalities of AST on transformer-based approach. 

Since those baselines are all applied to Java methods, we change their inputs as our smart contracts data, and follow their data processing steps to prepare their inputs. Besides, we adopt the greedy search algorithm in inference for all approaches to save the computation cost and ensure a fair comparison. 
Other noteworthy recent works, such as Wang \textit{et al}. \cite{wang2020reinforcement}, Zhang \textit{et al}.\cite{zhang2020retrieval}, adopted the Reinforcement Learning (RL) and Information Retrieval (IR) techniques based on recurrent models to improve the model  performance, respectively. 
Since the second improvement of the MMTrans focuses on the structural upgrade by the Transformer towards the recurrent models, the contribution of the Transformer-based structure towards the recurrent models cannot be distinguished from the RL and IR techniques based on recurrent models.
Therefore, the above two works are not suitable for comparison in this experiment. However, a potential future direction is to study the effect of the RL and IR techniques combining with the Transformer-based structure in another separate experiment.

\subsection{Metrics}
\label{metrics}
We adopt BiLingual Evaluation Understudy (BLEU) 
\footnote{NLTK: (\url{https://www.nltk.org/}) is used to calculate BLEU and METEOR\label{fn_metric}} \cite{papineni2002bleu}, 
Recall-Oriented Understudy for Gisting Evaluation (ROUGE) 
\footnote{rouge: (\url{https://github.com/pltrdy/rouge}) is used to calculate ROUGE} \cite{lin2004rouge},
and Metric for Evaluation of Translation with Explicit ORdering (METEOR) \textsuperscript{\ref{fn_metric}} \cite{denkowski2014meteor} to evaluate the performance of the code summarization approaches.
\begin{itemize}
\item  We report a composite BLEU score, which is the average of BLEU-1, BLEU-2, BLEU-3, and BLEU-4 (BLEU-n is the n-gram precision of a candidate sequence to the reference). Following Hu \textit{et al}.'s work \cite{hu2020deep}, we adopt both sentence-level BLEU (S-BLEU) and corpus-level BLEU (C-BLEU) to evaluate the generated comments, respectively. The S-BLEU calculates the composite BLEU score according to the sentence, and the C-BLEU calculates the composite BLEU score according to the whole corpus. In addition, to avoid non-overlapping n-grams in sentences, we simply use the smoothing-1 method \cite{chen2014systematic} to assist the computation of S-BLEU. 
\item ROUGE evaluates how much of reference text appears in the generated text, which can be thought of as a recall score. Following LeClair \textit{et al}.'s work \cite{10.1145/3387904.3389268}, we adopt the ROUGE-LCS (Longest Common Sub-sequence) F1 to evaluate the generated comments on the LCS matching degree between generated comments and references.
\item METEOR is also a widely used recall-oriented metric in machine translation and code summarization tasks. It evaluates how well the generated comments capture content from the references via recall, which is computed by stemming and synonymy matching. 
    
\end{itemize}
\subsection{Experimental Device}
The experiments are conducted on a Ubuntu GPU server with four RTX2080ti GPUs of 11 GB memory for each. Our proposed MMTrans is constructed by Tensorflow 2.3 based on CUDA 10.1 and cuDNN 7.4.

\begin{table*}[htbp]\caption{Automated Metrics Evaluation Results for the Baselines and MMTrans}\label{Automated Metrics Evaluation Results for the Baselines and MMTrans}
\centering
\begin{threeparttable}
\begin{tabular}{lccccc}
\toprule
Baseline & head number ($J$) & S-BLEU\tnote{*} (\%) & C-BLEU\tnote{*} (\%) & ROUGE-LCS F1 (\%) & METEOR (\%) \\
\midrule
Hybrid-DeepCom & / & 19.23 & 21.01 & 38.55 & 29.81 \\
code+gnn+GRU & / & 21.93 & 24.52 & 32.50 & 35.67 \\
Vanilla-Transformer & 4 & 25.99 & 28.79 & 47.36 & 39.25 \\
\midrule
Ours & & & & & \\
\midrule
i-MMTrans & 4 & 28.67 & 30.77 & 48.76 & 41.51 \\
MMTrans & 2 & 23.57 & 26.63 & 43.45 & 35.37 \\
MMTrans & 4 & \textbf{30.47} & \textbf{34.14} & \textbf{50.57} & \textbf{43.24} \\
MMTrans & 8 & 29.68 & 33.19 & 50.50 & 43.26 \\
MMTrans & 16 & 22.27 & 24.91 & 42.71 & 34.38 \\
MMTrans & 32 & 26.97 & 30.41 & 47.38 & 39.67\\
\bottomrule
\end{tabular}
\begin{tablenotes}
     \item[*] \small S-BLEU represents the Sentence-level BLEU score; C-BLEU represents the Corpus-level BLEU score.
\end{tablenotes}
\end{threeparttable}
\end{table*}

\section{RESULTS}
\label{experimental results}
This section reports the experimental results for the research questions proposed in Section \ref{Research Questions}.

\subsection{$RQ_{1}$: Quantitative Evaluation}
Table \ref{Automated Metrics Evaluation Results for the Baselines and MMTrans} shows the performance of the MMTrans ($J=4$) and the compared baselines. As shown in the table, the MMTrans performs the best, and achieves a S-BLEU score of 30.47, a C-BLEU score of 34.14, a ROUGE-LCS F1 score of 50.57, and a METEOR score of 43.24. 
Specifically, the MMTrans outperforms the Vanilla-Transformer by 17.23\% in terms of S-BLEU, by 20.74\% in terms of C-BLEU, by 6.78\% in terms of ROUGE-LCS F1, and by 10.17\% in terms of METEOR; the MMTrans also outperforms the code+gnn+GRU by 38.94\%, 39.23\%, 55.60\%, and 21.22\% in terms of the four metrics, respectively; and the MMTrans outperforms the Hybrid-DeepCom by 58.45\%, 62.49\%, 31.18\%, and 45.05\% in terms of the four metrics, respectively.
Towards the great improvement of the MMTrans against the three baselines, we generally attribute it to the multi-head attention structure and the utilization of both GCN and SBT to capture the local and global semantic information of code. We will discuss the strength of the MMTrans in more depth in Section \ref{strength of mmtrans}. 
We also find that the code+gnn+GRU outperforms the Hybrid-DeepCom in terms of S-BLEU, C-BLEU, and METEOR, while the latter performs better in terms of ROUGE-LCS F1.
This may owe to the higher attention of GCN on specific tokens and structures \cite{10.1145/3387904.3389268}, leading to the higher score in terms of the gram-statistic-oriented metrics (i.e., S-BLEU, C-BLEU, and METEOR).
However, SBT sequences represent the global semantic information of source code, therefore, the Hybrid-DeepCom can capture the global semantic information, and performs better in terms of the longest common sub-sequence matching (measured by ROUGE-LCS F1).
Moreover, even the Vanilla-Transformer only adopts the plain source code as the single input, its performance is still better than that of the Hybrid-DeepCom and code+gnn+GRU by a relatively large margin, which indicates the powerful capability of the Transformer model.

\textbf{Answer:} The MMTrans outperforms the Hybrid-DeepCom, code+gnn+GRU and Vanilla-Transformer by a significant margin in terms of all the automated evaluation metrics. 

\subsection{$RQ_{2}$: Head Number Analysis}
As mentioned in Section \ref{Research Questions}, the head number $J$ is the prominent part in the multi-head attention module compared with other attention structures, and affects the model performance to a great extent. In order to explore the influence of the $J$ in the MMTrans, we fix the other hyper-parameters and tune the $J = \{2, 4, 8, 16, 32\}$, respectively. The automated metric evaluation results are presented in Table \ref{Automated Metrics Evaluation Results for the Baselines and MMTrans}. The whole experiment result presents a general trend of increasing first then decreasing in terms of each of the evaluation metrics. And the performance reaches the peak when the $J = 4$. With the $J$ increasing, more representation subspaces of tokens appear but the embedding dimension of tokens decreases. As such, a potential explanation is that increasing the head number indeed can focus on more different perspectives; but when the subspaces proliferate excessively, the low representation dimension makes it impossible to accurately describe tokens, thus leading the attention distraction. Also notice that using head number $J = 32$ outperforms the $J=16$, this may due to the random initialization or other minor factors. 

\textbf{Answer:} When tuning the head number  $J = \{2, 4, 8, 16, 32\}$,
there is a general performance trend of increasing first then decreasing, and the MMTrans performs the best when the head number is 4.


\begin{table*}[htbp]\caption{Examples of Generated Comments by Each Approach}\label{Examples of Generated Comments by Each Approach}
\centering
\scriptsize
\begin{tabular}{l|p{70mm}|p{98mm}}

\toprule
ID & Smart Contract Methods & Comments \\
\midrule
\begin{tabular}[c]{@{}l@{}} 1 \end{tabular} & \begin{tabular}[c]{@{}l@{}} // Contract \#32527, Method \#7 \\ modifier whenCrowdsaleNotEnded \{\\ 	\quad	require(deadline \textgreater{}= now);\\ \quad		\_;\\ 	\}\end{tabular} & \begin{tabular}[c]{@{}l@{}}\textbf{Vanilla-Transformer:} modifier to make a function callable only when the contract is not finalized.\\ \textbf{i-MMTrans:} modifier to make a function callable only when the crowdsale has not ended.\\ \textbf{MMTrans: }modifier to allow actions only when the crowdsale is not ended.\\\textbf{Reference:} modifier to allow actions only when the crowdsale has not ended.\end{tabular} \\
\midrule
\begin{tabular}[c]{@{}l@{}} 2 \end{tabular} & \begin{tabular}[c]{@{}l@{}} // Contract \#37578, Method \#1 \\ function \_tokensToSell() \      private \     returns (uint256 tokensToSell) \{ \\  \quad    return latium.balanceOf(address(this));\\ \}\end{tabular} & \begin{tabular}[c]{@{}l@{}}  \textbf{Vanilla-Transformer:} function to sell tokens.\\ \textbf{i-MMTrans:} function to sell tokens ( with decimals ) that we are selling tokens.\\ \textbf{MMTrans:} function to get amount of Latium tokens ( with decimals ) of this contract.\\\textbf{Reference:} function to get current Latium balance of this contract.\end{tabular}\\
\midrule
\begin{tabular}[c]{@{}l@{}} 3 \end{tabular} & \begin{tabular}[c]{@{}l@{}}// Contract \#6743, Method \#27 \\ function finalize() public inState(State.Success) \\ \quad onlyOwner stopInEmergency \{\\     
\quad if(finalized) \ throw;\\     \quad if(address(finalizeAgent) != 0) \ finalizeAgent.finalizeCrowdsale();\\     \quad finalized = true;\\   \}\end{tabular} & \begin{tabular}[c]{@{}l@{}}\textbf{code+gnn+GRU:} finalize a succcesful crowdsale.\\ \textbf{i-MMTrans:} finalize a succcesful crowdsale.\\\textbf{Reference:} finalize a succcesful crowdsale.\end{tabular} \\
\midrule
\begin{tabular}[c]{@{}l@{}} 4 \end{tabular} & \begin{tabular}[c]{@{}l@{}} // Contract \#16118, Method \#5 \\ function playerMakeBet(uint minRollLimit, uint maxRollLimit, \\ \   bytes32 diceRollHash, uint8 v, bytes32 r, bytes32 s) public \   payable \\ \   gameIsActive  \   betIsValid(msg.value, minRollLimit, maxRollLimit)     \{\\  \quad      if (playerBetDiceRollHash{[}diceRollHash{]} != 0x0 \\ \quad \ $||$ diceRollHash == 0x0) throw;\\  \quad       tempBetHash = sha256(diceRollHash, byte(minRollLimit), \\ \quad \ byte(maxRollLimit), msg.sender);\\   \quad      if (casino != ecrecover(tempBetHash, v, r, s)) throw;\\   \quad  . . .\\  \quad       playerProfit{[}diceRollHash{]} = getProfit(msg.value, tempFullprofit);\\   \quad      if (playerProfit{[}diceRollHash{]} $>$ maxProfit)              throw;\\       \quad  . . .\\   \quad      LogBet(diceRollHash, playerAddress{[}diceRollHash{]}, \\ \quad \ playerProfit{[}diceRollHash{]},     playerToJackpot{[}diceRollHash{]}, \\ \quad \ playerBetValue{[}diceRollHash{]},  playerMinRollLimit{[}diceRollHash{]}, \\ \quad \ playerMaxRollLimit{[}diceRollHash{]});\\     \}\end{tabular} & \begin{tabular}[c]{@{}l@{}}\textbf{code+gnn+GRU: }appends the bid's.\\ \textbf{i-MMTrans:} public function player submit bet only if game is active bet is valid can be called.\\ \textbf{Reference:} public function player submit bet only if game is active bet is valid.\end{tabular}\\
\bottomrule
\end{tabular}
\end{table*}

\subsection{$RQ_{3}$: Strength of the MMTrans}
\label{strength of mmtrans}
This section elaborates on the strength of MMTrans from two perspectives, i.e., heterogeneous multi-modalities of AST and the multi-head attention structure.

(1) The Vanilla-Transformer uses the plain source code as a single input; the i-MMTrans has double inputs of graphs and plain source code, while the MMTrans has double inputs of graphs and SBT sequences. Moreover, these approaches all employ the multi-head attention structure. As shown in Table \ref{Automated Metrics Evaluation Results for the Baselines and MMTrans}, we find that the i-MMTrans outperforms the Vanilla-Transformer by 10.31\% in terms of S-BLEU, by 6.88\% in terms of C-BLEU, by 2.96\% in terms of ROUGE-LCS F1, and by 5.76\% in terms of METEOR, which indicates that the additional input of graphs indeed boosts the model performance. Subsequently, we make a comparison between the i-MMTrans and the MMTrans. The Table \ref{Automated Metrics Evaluation Results for the Baselines and MMTrans} demonstrates that the latter outperforms the former by 6.28\%, 10.95\%, 3.71\%, and 4.17\% in terms of the four metrics respectively, which proves that the additional input by SBT sequences can further improve the model performance. 

Meanwhile, we present the first two instances to intuitively show the strength of the heterogeneous multi-modalities of AST in Table \ref{Examples of Generated Comments by Each Approach}. 
Following LeClair \textit{et al.}'s work \cite{10.1145/3387904.3389268}, we intuitively illustrate the power of multi-modalities of AST from the copy mechanism perspective. For the first instance, the i-MMTrans can directly copy the correct words, such as ``modifier'', ``when'', ``crowdsale'' and ``ended'', from the source code to the comment. Comparing with the Vanilla-Transformer, its copy capability is indeed more powerful. However, the i-MMTrans seems relatively weaker in summarizing the main idea of source code, when the incorrect words accounts for a relatively large ratio. Noticing the second instance, the i-MMTrans wrongly copies the word ``sell'' and ``tokens'', which appear with the highest frequency in the source code, therefore does not capture the main idea of the code snippet. A potential explanation is the i-MMTrans equipped with GCN puts more focus on local features of the AST, causing it easy to be attracted by the specific words and code structure. However, the MMTrans equipped with both the GCN and SBT can weigh between local and global semantic information, therefore modifies the fault caused by the i-MMTrans, and catches the correct main idea of the source code in the second instance.

(2) As mentioned in Section \ref{Methodology}, the i-MMTrans and code+gnn+GRU both exploit the graph and plain source code as the double inputs, but the former adopts the multi-head attention structure while the latter adopts the GRU-based structure. Statistically, the i-MMTrans outperforms the code+gnn+GRU by 30.73\% in terms of S-BLEU, by 25.49\% in terms of C-BLEU, by 50.03\% in terms of ROUGE-LCS F1, and by 16.37\% in terms of METEOR, which indicates that applying the multi-head attention structure indeed improves the performance a lot in the source code summarization task. And it is noticeable that the contribution to the performance improvement of the multi-head attention structure is greater than that of the multi-modalities of AST. We also list two examples of the generated comments in the Table \ref{Examples of Generated Comments by Each Approach}. The instance \#3 is a short method that can be parsed to the plain source code with 49 tokens and the graph node sequence with 59 tokens. The i-MMTrans and code+gnn+GRU can both generate the exact correct comments on this short sample. Nevertheless, when it comes to a relatively long method, such as the instance \#4 with 152 tokens in the plain source code and 200 tokens in the graph node sequence (both have been truncated to the maximum length of the above two sequences), the code+gnn+GRU with the GRU-based structure cannot summarize the correct comment, while the i-MMTrans can still generate readable and meaningful comment. The reason is that the multi-head attention structure can properly allocate different attention weights on the tokens at each time step, and summarize their key information for inference; however, the GRU-based network has a limited capability in capturing the long-range dependencies between code tokens, thus generating the low-quality comments for relatively long methods. 

\textbf{Answer: } Leveraging SBT sequences and graphs to extract the global and local semantic information of code, and employing the  multi-head attention structure to capture the long-range dependencies between code tokens contribute to the generation of higher-quality comments. 

\section{THREATS TO VALIDITY}
\label{threats}
We have identified the following threats to validity:

\noindent \textbf{Dataset quality:} As the first smart contract code summarization work, we used some heuristic rules to extract the $<$method, comment$>$ pairs according to the characteristic of smart contract data. Although we did a rigorous data processing, there may be still some noise. We will continuously refine and update the version of the open-source dataset.
\noindent \textbf{Comparison on smart contract dataset:} Since our work focuses on the smart contract code summarization, we did not compare the MMTrans with Hybrid-DeepCom, Vanilla-Transformer, and code+gnn+GRU on their datasets. But the results on smart contracts also have proved the effectiveness of the MMTrans. In the future, we will extend our experiment on other programming languages (e.g., Java and Python).

\noindent \textbf{Fair comparison threat:} Due to the hardware limitations, we were unable to conduct fully extensive hyper-parameters optimization for all baselines. Following LeClair \textit{et al}.'s work \cite{10.1145/3387904.3389268}, we try to mitigate the impact of this issue by making each baseline's hyper-parameters consistent with our model. 

\noindent \textbf{Automated evaluation:} We adopt four automated evaluation metrics that have been widely used in previous code summarization studies \cite{hu2018deep,10.1145/3387904.3389268,hu2020deep,ahmad-etal-2020-transformer}.
Although the metrics are not representative of human judgment \cite{stapleton2020human}, they can evaluate the performance of code summarization models quickly and quantitatively. In the future, we will conduct human evaluation on the models.

\section{CONCLUSION}
\label{conclusion}
This work aims to help programmers comprehend the meaning of smart contract code by automatically generating high-quality comment. To tackle this task, we for the first time collect a smart contract code summarization dataset with 347,410 $<$method, comment$>$ pairs. Meanwhile, we propose a code summarization approach named MMTrans, which leverages the two modalities of the AST (i.e., SBT sequences and graphs) to represent both global and local semantic information of source code, then employs the two encoders and a joint decoder with the multi-head attention structure to capture the long-range dependencies between code tokens. The comprehensive experiments on the collected dataset show that the MMTrans performs better than the state-of-the-art baselines by a significant margin, and can generate higher-quality comments in the practical tests.

\section*{Acknowledgment}
This work is supported in part by the General Research Fund of the Research Grants Council of Hong Kong (No. 11208017) and the research funds of City University of Hong Kong (7005028 and7005217), and the Research Support Fund by Intel (9220097), and funding supports from other industry partners (9678149,9440227, 9229029, 9440180 and 9220103).





\bibliographystyle{IEEEtran}
\bibliography{ref}{}

\end{document}